# High-$T_C$ Superconductivity in Hydrogen Clathrates Mediated by Coulomb Interactions Between Hydrogen and Central-Atom Electrons

Dale R. Harshman [1] · Anthony T. Fiory [2]

**Abstract**

The uniquely characteristic macrostructures of binary hydrogen-clathrate compounds $M$H$_n$ formed at high pressure, a cage of hydrogens surrounding a central-atom host, is theoretically predicted in various studies to include structurally stable phonon-mediated superconductors. High superconductive transition temperatures $T_C$ have thus far been measured for syntheses with $M$ = La, Y, and Th. In compressed LaH$_{10}$, independent studies report $T_C$ of 250 K and over 260 K, a maximum in $T_C$ with pressure $P$, and normal-state resistance scaling with temperature (suggesting unconventional pairing). According to reported band structure calculations of $Fm\bar{3}m$-phase LaH$_{10}$, the La is anionic, with the chemical valence electrons appearing evenly split between La and H$_{10}$. Thus, compressed LaH$_{10}$ contains the combination of structure, charge separation and optimal balanced allocation of valence electrons for supporting unconventional high-$T_C$ superconductivity mediated by Coulomb interactions between electronic charges associated with La and H$_{10}$. A general expression for the optimal superconducting transition temperature for $M$H$_n$ clathrates is derived as $T_{C0} = k_B^{-1}\Lambda[(n + v)/2A]^{1/2}e^2/\zeta$, where $\Lambda$ is a universal constant, $(n + v)$ is the chemical valence sum per formula unit, taking unity for H and $v$ for atom $M$, $A$ is the surface area of the H-polyhedron cage, and $\zeta$ is the mean distance between the $M$ site and the centroids of the polyhedron faces. Applied to $Fm\bar{3}m$ LaH$_{10}$, $T_{C0}$ values of 249.8(1.3) K and 260.7(2.0) K are found for the two experiments. Associated attributes of charge allocation, structure, effective Coulomb potential, and H-D isotope effect in $T_C$ of $Fm\bar{3}m$ LaH$_{10}$ and $Im\bar{3}m$ H$_3$S are discussed, along with a generalized prospective for Coulomb-mediated superconductivity in $M$H$_n$.

Dale R. Harshman
drh@physikon.net

[1] Department of Physics, The College of William and Mary, Williamsburg, VA 23187, USA
[2] Department of Physics, New Jersey Institute of Technology, Newark, NJ 07102, USA

# 1 Introduction

Binary hydride clathrates $M$H$_n$ are a class of compounds comprising a central atom surrounded by a polyhedron hydrogen cage and have been designated as superhydrides. Following earlier theoretical results [1], superconductivity at high temperatures has been predicted in dozens of such (and other) hydride compounds under compression, assuming strong-coupled electron-phonon theory [2-4]. Notably, the superconductors LaH$_{10}$ [5, 6], YH$_n$ (n = 4, 6, 9) [5-9], and ThH$_n$ (n = 9, 10) [6, 10] have been experimentally verified in [11-13], [14, 15], and [16], respectively. The LaH$_{10}$ clathrate, exhibiting superconductivity at temperatures of about 250 K [12] and above 260 K [13] for applied pressures $P \sim 137 - 218$ GPa, currently holds the record for the highest transition temperature, $T_C$. Electron-phonon interaction theory rooted in strong-coupling Migdal-Éliashberg methodology [17-19], density functional theory, or *ab initio* methods [4, 20], predicts $T_C$ for LaH$_{10}$ spanning a broad range of 199 – 288 K [4-6, 21-26] and monotonically decreasing with increasing pressure [5, 24-26]. The $Fm\bar{3}m$ phase of LaH$_{10}$ is theoretically stable at $P = 129 - 264$ GPa [25].

Although electrical resistance measurements are generally employed as the principal indicator of superconductivity, the reported data already indicate unconventional behavior, particularly for LaH$_{10}$ and YH$_9$. For each compound, a maximum $T_C$ is observed in its pressure dependence, referred to as a "dome" in $T_C$-*vs*.-*P* [12, 14], setting bounds on pressure for maximum $T_C$ as found in unconventional high-$T_C$ superconductors, such as in the compressed phases of Cs$_3$C$_{60}$ [27]. Differences between the LaH$_{10}$ samples reported in [12] and [13] indicate a 10-K increase in $T_C$ related to increased pressure, similar to the influence of higher synthesis pressure noted in [28] and the pressure dependence reported in [29]. Negative pressure shifts in $T_C$ calculated in [5, 23-26] evidently do not capture these experimental findings. In the normal state, the electrical resistance of LaH$_{10}$ and YH$_9$ samples is found to be nearly proportional to temperature [12-14, 30], exhibiting the characteristic absence of resistance saturation of unconventional high-$T_C$ superconductors [31].

A review of conventional electron-phonon coupling theory in [4] includes a listing of numerous computational flaws that remain to be addressed.[1] As independent corroboration of phonon mediation has yet to be convincingly established [36], alternative considerations of electronic mechanisms are being developed, particularly in view of the high pressures [37], correspondences to high-$T_C$ cuprates [38] and limitations on the Fermi temperatures [39].

An unconventional Coulomb-mediated high-$T_C$ mechanism, wherein superconductivity originates locally within the metal-clathrate structure, is proposed herein which is independent of conventional phononic pairing involving the extended crystalline lattice. The proposed model is a novel adaptation of the inter-reservoir Coulomb-mediated superconducting pairing theory, first introduced in [40], and successfully applied to compressed $Im\bar{3}m$ H$_3$S [41],[2] which asserts that quasiparticle pairing arises from Coulomb interactions between charges in two adjacent, spatially-separated reservoirs, designated as type I and type II. In the case of compressed $LaH_{10}$, the hydrogen polyhedron cage is identified as type I (superconducting) and the central La as type II (mediating). These charge reservoirs are separated by an interaction distance $\zeta$ determined as the mean distance between the polyhedron center and the centroids of its faces. The requisite balance between type I and type II charges for optimal high-$T_C$ superconductivity posited in [40] appears satisfied by the anionic nature of the La charge as revealed by theoretical band structure of $LaH_{10}$ at high pressure [5, 22]. As shown in Section 2, each reservoir contains a number of charges per formula unit $\sigma$ equal to one-half of the total number of chemical valence charges, which is optimally 13 for $LaH_{10}$.

---

[1] The layered FeH$_5$, synthesized with no demonstration of superconductivity [32], is predicted to be superconducting in [33, 34] and non-superconducting in [35].

[2] Note that here is a typographical error at the end of Section 3 (p. 5) of [41] which should read, "$\ell = (A/\sigma)^{1/2} = 2.883$ Å," with the reported $T_{C0}$ remaining unchanged.

In Section 2, a general expression for $T_{C0}$ is derived for $MH_n$ clathrates and applied to calculate $T_{C0}$ of $LaH_{10}$. Section 2 also considers the question of superconductivity and charge equilibrium between the two reservoirs in $LaH_{10}$, $YH_n$, and $ThH_n$, expressing $T_{C0}$ as a function of the average H-H bond length in the binary clathrates. Section 3 considers potential superconducting clathrates in terms of inter-reservoir charge equilibrium and phase purity. Also, data for the isotope effect in $T_C$ reported in [12] are analyzed from the perspective of sample-quality variations, similar to that done for $Im\bar{3}m$ $H_3S$ [42]. Finally, implications of the Coulomb potential signature in optical reflectivity data and relevant electron-phonon coupling in $LaH_{10}$, are considered. Results are summarized and conclusions drawn in Section 4. The Appendix contains theoretical and experimental $T_C$-vs-$P$ results for $LaH_{10}$, $YH_9$, and $YH_6$ (Section A1), normal-state resistance (Section A2), pressure dependence of the $LaH_{10}$ lattice parameter (Section A3), and accuracy and systematic errors of the interlayer Coulombic pairing model (Section A4).

## 2 Coulomb-Mediated Superconductivity in $MH_n$

Experimental bases for unconventional high-$T_C$ superconductivity and optimal $T_C$ are the non-monotonic "dome"-shaped pressure dependence in $T_C$ reported for $Fm\bar{3}m$ $LaH_{10}$ [12] and $P6_3/mmc$ $YH_9$ [14]. This signature feature of high-$T_C$ materials defines optimal $P$ and $T_C$ at the top of the dome where $T_C$ is highest [31]. As shown by the samples numbered 1 – 6 in [12], $T_C$ systematically increases with increasing $P$ at sub-optimal pressures. Conventional electron-phonon interaction theory predicts a fundamentally different monotonic decrease in $T_C$ with increasing pressure above a structural stability threshold [24-26], and resembling the behavior of depressed $T_C$ at pressures above optimal [12, 13]. Values of optimal $T_C$ and corresponding pressures are determined from data for $LaH_{10}$ [12, 13] and $YH_9$ [14] (Section A1).

Unconventional high-$T_C$ superconductivity in the hydrogen clathrates is further demonstrated in the temperature dependence of the normal-state resistance, $R_N(T)$. As is discussed in Section A2, $R_N(T)$ for both $Fm\bar{3}m$ $LaH_{10}$ [12, 13] and $P6_3/mmc$ $YH_9$ [14] is linear, exhibiting an absence of saturation at high temperature, indicative of an absence of strong electron-phonon coupling, and a near-zero intercept extrapolation to zero temperature, $R_N(T\rightarrow 0) \sim 0$. Given that both materials exhibit this unconventional behavior, it is unlikely coincidental. Recall that the lack of saturation and $R_N(T) \propto T$ are well-recognized features of optimal high-$T_C$ materials [31, 43], strongly suggesting a pairing mechanism akin to that of other (unconventional) high-$T_C$ superconductors. In optimally- and over-doped high-$T_C$ cuprates, $R_N(T) \propto T$ is generally attributable to "Planckian" dissipation [44, 45]. Analysis of $R_N$ in Section A2 indicates Planckian dissipation in LaH$_{10}$ comparable to the high-$T_C$ cuprates [46-48].

Superconductive pairing in high-$T_C$ superconductors is successfully modeled by Coulomb interactions between charges in spatially separated charge reservoirs [40, 41, 49-55]. Developed initially to describe indirect Coulombic pairing interactions between charges in adjacent extended layers, this unique theory is also applicable to extended three-dimensional (3D) lattices, such as the compressed phases of Cs$_3$C$_{60}$ [49] and H$_3$S [41], in which charges are similarly separated into two reservoirs. To date, this approach has been validated with statistical accuracy of ±1.30 K (or ±4%) in $T_{C0}$ for 53 different materials from ten disparate superconducting families; the aforecited 3D compounds [41, 49], layered cuprates, ruthenates, rutheno-cuprates, iron pnictides, BEDT-based [bis(ethylenedithio)tetrathiaful-valene] organics [40, 50, 51], iron chalcogenides [52], intercalated group-4-metal nitride halides [53, 54] and gated twisted bilayer graphene [55], with measured $T_{C0}$ values ranging from ~2 to 200 K. Optimization resulting in a maximum transition temperature is typically accomplished via doping and/or pressure-induced charge transfer.

In general terms, the type I reservoir is identified with the dominant superconducting condensate while the type II reservoir is associated with mediating charges. However, in some instances, both reservoirs serve dual superconducting and mediating purposes [41, 53]. Given the electronic and structural features required as described, the model predicts an optimal transition temperature $T_{C0}$ of a homogeneous high-$T_C$ superconductor given by the algebraic formula scaling with the potential energy $e^2/\zeta$ [40],

$$T_{C0} = k_B^{-1} (\Lambda/\ell)\, e^2/\zeta, \qquad (1)$$

where $\zeta$ is the perpendicular distance between the two interacting charge reservoirs, and the length $\ell = (\sigma\eta/A)^{-1/2}$ is the linear interaction charge spacing defined by the participating charge $\sigma$ (a dimensionless fraction) per unit area $A$ of the type I reservoir. The factor $\eta$ is the number of type II mediating layers in certain topographies with multiple layers and equals unity for the binary superhydride clathrates. The constant $\Lambda = 0.007465(22)$ Å $= 1.933(6)\lambdabar_C$ is previously determined experimentally to a ±0.3% accuracy in [40, 41] and expressed in terms of the reduced Compton wavelength $\lambdabar_C$. The microscopic description leading to Eq. (1) is discussed in [41].

For the superhydride clathrates $M$H$_n$, the type I charge reservoir structure is identified as the cage of hydrogen ions, each located at the vertices of a polyhedron, and the type II charge reservoir is the enclosed central metal ion $M$. The polyhedron is treated as a closed layer that is folded upon itself in 3D, and thus its surface area determines $A$ for the type I reservoir. The average of the distances $\zeta_i$ between the polyhedron center and its polygon-face centers, weighted by their areas $A_i$, determines $\zeta$ according to

$$\zeta = \Sigma_i\, \zeta_i A_i/A\,. \qquad (2)$$

These definitions of $A$ and $\zeta$ are drawn from earlier work on Cs$_3$C$_{60}$, following methods initially established in [40] for extended layered structures. The hydrogen cage provides the same type I function as the C$_{60}$ in compressed Cs$_3$C$_{60}$, wherein the corresponding type II reservoir comprises nearest-neighbor cesium ions located outside of the C$_{60}$ structure [49].

The total charge per formula unit forming the superconductive condensate [40] is created for some materials by doping through atomic substitutions in one or both charge reservoirs [52] while it is inbuilt in others by available chemical valence electrons. An example of the latter is the three valence electrons from Cs per formula unit of $Cs_3C_{60}$ [49]. Here, the participating charge σ is determined by multiplying the total number of electron charges by the inter-reservoir sharing factor γ, which follows the charge allocation rules in [40] (see also [51]). For the $MH_n$ clathrates only rule (1b) applies,[3] specifying γ = 1/2 for equal sharing of the total charge between the two reservoirs. The total charge number equals the sum of the full chemical valence per formula unit of $M$ with valence $v$ and the n hydrogens, each with unity valence. Whether the charge (n + $v$) is equally shared between $M$ and $H_n$ ions, as is found for $H_3S$ [41], remains to be determined by examining the particular electronic structure of each $MH_n$ compound. For optimal charge allocation between the two reservoirs, γ = 1/2 and the resulting equation for the participating charge

$$\sigma = (n + v)/2 , \qquad (3)$$

thus determines the length $\ell$ in Eq. (1),

$$\ell = [(n + v)/2A]^{-1/2} , \qquad (4)$$

which then becomes

$$T_{C0} = k_B^{-1}\Lambda[(n + v)/2A]^{1/2}e^2/\zeta . \qquad (5)$$

Optimization in $T_C$ is achieved via hydrostatic compression, suggesting a charge transfer process which induces equilibrium between the two charge reservoirs. Making correspondence to the prescription of rule (1b) [40], the $M$ ion (type II) acquires (relative to the neutral atom) the optimal number (σ − $v$) = (n − $v$)/2 of electrons and the n H ions (comprising type I) collectively give up the number (n − σ) = (n − $v$)/2 of electrons, equivalent to forming holes of like number.

The concept of charge equilibrium between the interacting structures of the two reservoirs is an important requirement for achieving an optimal high-$T_C$ superconductive state within the interlayer Coulomb mediation model. It specifies equality of reservoir charges σ in 3D structures such as $Cs_3C_{60}$ [49] and the binary clathrates $MH_n$, which is equivalent to equal areal charge densities $\ell^{-2}$ for interacting layers in 2D layered reservoir structures as derived in [40]. One thus expects adherence to charge equality for optimal high-$T_C$ superconductivity in the binary super-hydrides. Charge equilibrium between the two reservoirs, which is not explicitly considered in electron-phonon theory, dictates both the presence and quality of the superconductivity in $MH_n$ clathrates. In contrast, superconducting behavior in electron-phonon theory is determined by the dense hydrogen-clathrate sublattice [26] and the mass [6], rather than the chemical details [8] of the enclosed atom. In the following, structure, charge allocation, and optimal $T_{C0}$ are determined for $LaH_{10}$ (Section 2.1), $MH_n$ generally (Section 2.2), and other clathrates such as $YH_n$ (Section 2.3).

## 2.1 $LaH_{10}$ – Calculation of $T_{C0}$

From Eq. (3), the participating charge fraction σ = (10 + 3)/2 = 6.5 is obtained by adding up the valence electrons in the atoms constituting a formula unit of $LaH_{10}$, counting n = 10 for the ten H $1s^1$ and $v$ = 3 for a single La $5d^16s^2$. Although 6.5 exceeds the normal number of 3 valance electrons in atomic La, band structure calculations of $Fm\bar{3}m$ $LaH_{10}$ at $P$ = 300 GPa show formation of hybridized La-$f$ and H-$s$ orbitals, rendering an anionic character to the La [22]. An electron density contour map presented in Fig. 1(b) of [22] also shows that the La anion is separated from its neighboring hydrogen ions by a shell of comparatively reduced electron density of approximately 0.5 – 0.6 Å$^{-3}$. By defining boundaries between ions along such minima in local electron density, one estimates that about 46 ± 4% of the total valence electronic charge is on the anionic La, with 44 ± 4% at the eight H1 ions (also designated anionic in [22]) and 10 ± 1% at the two H2 ions (designated cationic in [22]); the total number of electrons estimated in this manner is 14 ± 1 ≈ (n + v) per formula unit. Hence, this analysis of results from band structure calculations suggests that the total valence charge is nearly equally divided between the La and the ten H ions, affirming application of Eq. (3) and its derivation from rule (1b) [40].

---
[3] Rule (1b) from [40] (see also [51]) reads that "The doping is shared equally between the hole and electron reservoirs, resulting in a factor of 1/2."

The anionic character of La is also indicated by the partial density of electron states (PDOS) for LaH$_{10}$ at 250 GPa in occupied *s*, *p*, *d*, and *f* orbitals projected onto La, as shown in Fig. S3(b) of [5]. The electron charge on H$_{10}$ is determined by integrating the PDOS shown for orbitals projected onto H. Over the plotted range −1.5 Ry ≤ $E$ ≤ $E_F$, the H-projected *s* and *p* orbitals contain charges 5.7$e$ and 0.7$e$, respectively, giving a total of at least 6.4$e$ on H$_{10}$. Integrated La-projected PDOS is at least 15.5$e$ and evidently includes semi-core states. Unplotted PDOS for $E$ < −1.5 Ry might slightly increase these numbers. The number of electron charges in the type I H$_{10}$ reservoir at 250 GPa is evidently within 1.5% from being equal to the number σ = 6.5 per formula unit determined from rule (1b) and calculated according to Eq. (3).

In the $Fm\bar{3}m$ structure of LaH$_{10}$ the type I polyhedron cage comprises 32 hydrogens at the vertices of a chamfered cube with square and irregular hexagon faces (see Fig. 1). The 24 type H1 ions at vertices of adjoining square and hexagon faces are shared among 3 polyhedra and the 8 type H2 ions at vertices of adjoining hexagon faces are shared among 4 polyhedra, allocating 24/3 = 8 H1 ions and 8/4 = 2 H2 ions, respectively, per formula unit. Structure theory has determined irregular chamfered cube geometry with H1-H1 bonds longer than H1-H2 bonds [5, 22].

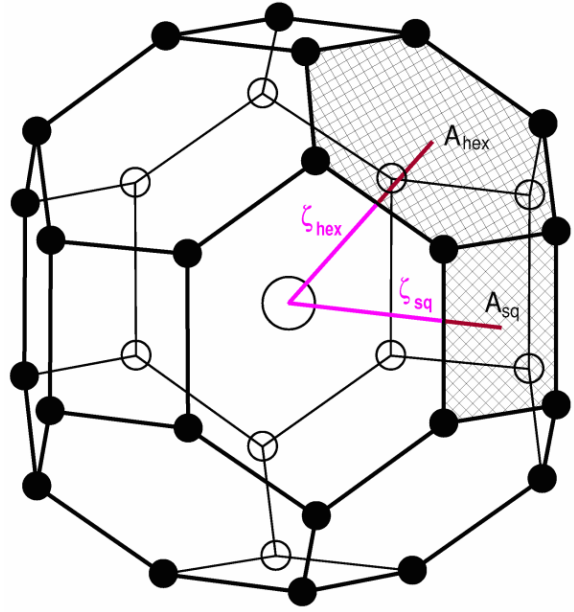

**Fig. 1** The LaH$_{10}$ clathrate structure, with hydrogens (small spheres) at the vertices of a chamfered cube polyhedron and La (large sphere) at center. The two interaction lengths, $\zeta_{hex}$ and $\zeta_{sq}$ extend from the center to the centroids of the hexagonal and square polyhedron faces of area $A_{hex}$ and $A_{sq}$, respectively

Recognizing that the hydrogen ion sites and occupancies have not been directly determined from x-ray diffraction of $Fm\bar{3}m$ LaH$_{10}$, the quantities in Eq. (2) are calculated using theoretical hydrogen bond lengths in irregular chamfered cube structures [5, 22], in combination with measurements of the lattice parameter *a* [12, 13]. Defining the ratio of the H1-H1 to H1-H2 bond lengths as $r$ = $d_{H1-}$ $_{H1}/d_{H1-H2}$ and the associated anisotropy factor $f = (r − 1)/(r + 1)$, the average H-H bond length, expressed in terms of lattice parameter, is $d = a[2 + 4/3^{1/2} + f(2 − 4/3^{1/2})]^{-1}$. In the irregular chamfered cube polyhedron, four H1-H1 bonds form 6 square faces of area $A_{sq} = (rd)^2$, two H1-H1 and four H1-H2 bonds form 12 irregular hexagon faces of area $A_{hex} = [8·2^{1/2}(1/3 + r/3^{1/2})/(1 + r)^2]d^2$, and the total area of the polyhedron is thus $A = 6A_{sq} + 12A_{hex}$. Distances measured from polyhedron center are $\zeta_{sq} = d[(r + 4/3^{1/2})/(1 + r)]$ to the square face centers and $\zeta_{hex} = d[2^{-1/2}(r + 2/3^{1/2})/(1 + r)]$ to the hexagon face centers; the average distance weighted by face areas gives $\zeta = (6A_{sq}\zeta_{sq} + 12A_{hex}\zeta_{hex})/A$. These distance and area parameters are illustrated in Fig. 1. Below, the expressions for $d$, $A$, and $\zeta$ are evaluated for $r = 1.077$, which is obtained by averaging the theoretical pressure dependence of $d_{H1-H1}$ and $d_{H1-H2}$ calculated in [5] over the optimal range $P = 169 –192$ GPa determined from experiment (Section A1).

Data presented in the inset of Fig. 1 of [12] for $Fm\bar{3}m$ LaH$_{10}$ show the highest superconducting transition temperature occurring for sample 5 at $T_C = 251(1)$ K at a corresponding average pressure $P = 169(4)$ GPa (Section A1). The corresponding lattice parameter $a = 5.009(14)$ Å is determined by the variation of $a$-vs.-$P$ derived from the experimental data (Section A3). The formulae given above then evaluate to $d = 1.165(3)$ Å, $A = 50.29(27)$ Å$^2$, $\zeta = 1.795(5)$ Å, $\ell = 2.781(8)$ Å, and $T_{C0} = 249.8(1.3)$ K. These results are listed in the first data row of Table 1.

A complete superconducting resistance transition is shown in Fig. 1 of [13] for a sample at initial pressure of 188 GPa, where the authors observe an appreciable drop in resistance at 260 K upon cooling (blue cooling curve) and a pressure of 196 GPa after warming back to 300 K. As discussed in Section A1, the superconducting onset $T_C = 262(1)$ K is considered to be optimal at the average pressure of $P = 192(4)$ GPa. The corresponding lattice parameter is $a = 4.903(18)$ Å. One thereby obtains for this sample the values $d = 1.141(4)$ Å, $A = 48.18(36)$ Å$^2$, $\zeta = 1.757(7)$ Å, $\ell = 2.723(10)$ Å, and calculated $T_{C0} = 260.7(2.0)$ K, which are included the second data row of Table 1.

Third row entries in Table 1 describe results for compressed $Im\bar{3}m$ H$_3$S. The measurement of $T_C$ is taken from the most recently available data, identified as "155 GPa: 11/2018" in Fig. 1 of [56], exhibiting a significantly sharper transition than that reported previously [57]. In [41] showing Coulomb-mediated superconductive pairing in $H_3S$, a charge of $\sigma = 3.43 \pm 0.10$ is derived from the average of integrated PDOS for H $s$ electrons calculated in [58] and [59] (per $H_3$ unit). Much of the PDOS for the sulfur $s$ orbital lies deep in energy, diminishing near $E_F$, as shown in Fig. 2b of [58] and in Fig. 2 of [60]. Evidently, the two $3s^2$ electrons from atomic S fill localized core-like orbitals such that the available valence electrons are dominated by the four $3p^4$ electrons of the S atom and the three $1s$ electrons from $H_3$. In reference to the value of $\sigma$ derived from projected H $s$ orbitals, inter-reservoir charge equilibrium is evidently established with S valence $\nu = (3.86 \pm 0.20) \approx 4$ and n = 3 in Eq. (3) and produced by transfer of $(0.47 \pm 0.10)e \approx e/2$ valence electron charge from S to $H_3$. Model parameters determined previously [41] are shown for comparison, but in this case using Eq. (3) with n = 3 and $\nu = 4$ to give $\sigma = 7/2 = 3.5$ and $T_{C0} = 200.6(3)$ K.

**Table 1** Experimental transition temperatures $T_C$ at applied pressures $P$ are given for $Fm\bar{3}m$ LaH$_{10}$ from data in [12, 13] and for $Im\bar{3}m$ H$_3$S from data in [56]. Lattice parameter $a$ is determined by $P$. Theoretical parameters listed are the participating charge σ, average H–H bond length $d$, surface area $A$ (H$_{32}$ polyhedron for $Fm\bar{3}m$ LaH$_{10}$, cubic faces for $Im\bar{3}m$ H$_3$S [41]), interaction distance $\zeta$, mean spacing between participating charges $\ell$ and calculated optimal transition temperature $T_{C0}$, as defined in the text

| Expt. Ref. | $T_C$ (K) | $P$ (GPa) | $a$ (Å) | σ | $d$ (Å) | $A$ (Å$^2$) | $\zeta$ (Å) | $\ell$ (Å) | $T_{C0}$ (K) |
|---|---|---|---|---|---|---|---|---|---|
| LaH$_{10}$ ($Fm\bar{3}m$) | | | | | | | | | |
| [12] | 251(1) | 169(4) | 5.009(14) | 6.5 | 1.165(3) | 50.29(27) | 1.795(5) | 2.781(8) | 249.8(1.3) |
| [13] | 262(1) | 192(4) | 4.903(18) | 6.5 | 1.141(4) | 48.18(36) | 1.757(7) | 2.723(10) | 260.7(2.0) |
| H$_3$S ($Im\bar{3}m$) | | | | | | | | | |
| [56] | 201 | 155 | 3.082(5) | 3.5 | 2.180(4)[a] | 28.50(9) | 2.180(4) | 2.854(5) | 200.6(3) |

[a] $d$ is intra-sublattice H-H distance

Figure 2 shows the experimental values of $T_C$ vs. $(\ell\zeta)^{-1}$ from the preceding calculations for $Fm\bar{3}m$ $LaH_{10}$ and given in Table 1. Horizontal error bars correspond to uncertainties in pressure; uncertainties in $T_C$, smaller than symbols sizes, are not shown. For comparison, the corresponding result for $H_3S$ [56] from Table 1 is shown by the triangle symbol. The solid line in Fig. 2 represents theoretical dependence of $T_{C0}$ on $(\ell\zeta)^{-1}$ defined by Eq. (1). Given the essential differences in structure and bonding between $Fm\bar{3}m$ $LaH_{10}$ and $Im\bar{3}m$ $H_3S$ (noted in [61]), it is remarkable that theory derived from the interlayer Coulomb-mediation pairing model accurately determines $T_{C0}$ for both hydride families. A comparison of the two optimal $LaH_{10}$ clathrate samples with $H_3S$ and optimal materials from other unconventional high-$T_C$ families is given in Section A4.

## 2.2 Optimal $MH_n$ Clathrates

With the proviso that the optimization criterion could satisfied for some binary $MH_n$ clathrates other than $LaH_{10}$, it is useful to note that Eqs. (2) − (5) have a general functional dependence on the participating charge $\sigma$ and average H-H bond length $d$ in the form

$$T_{C0} = C_n \sigma^{1/2} k_B^{-1} \Lambda e^2 / d^2 , \qquad (6)$$

where $C_n = A^{-1/2}\zeta^{-1}d^2$ depends only on the H-cage structure (cubic symmetries have $d \propto a$ and $T_{C0} \propto a^{-2}$). Values of $A/d^2$, $\zeta/d$, and $C_n$ are given in Table 2 for exemplary $Fm\bar{3}m$ $YH_3$, $Im\bar{3}m$ $YH_6$, $P6_3/mmc$ $YH_9$ and $Fm\bar{3}m$ $YH_{10}$ clathrates corresponding to n = 3, 6, 9 and 10 [6, 8]. The corresponding theoretical crystal symmetries are H-cage polyhedral structures, $[4^{12}]$ (rhombic dodecahedron), $[4^6 6^8]$ (truncated octahedron), $[4^6 5^6 6^6]$ (irregular polygon faces) and $[4^6 6^{12}]$ (irregular

**Table 2** Geometric factors for hydrogen-cage polyhedra in $MH_n$ with specified structures. Polyhedron surface area $A$ and average distance $\zeta$ between the polyhedron center and its face-polygon centroids, weighted by area, are given normalized to average polyhedron edge length $d$, which determine the numerical coefficient $C_n = A^{-1/2}\zeta^{-1}d^2$ in Eq. (6)

| n | Structure | Vertex number | $A/d^2$ | $\zeta/d$ | $C_n$ |
|---|---|---|---|---|---|
| 3 | $[4^{12}]$ [a] | 14 | 11.314 | 0.8165 | 0.3641 |
| 6 | $[4^6 6^8]$ [b] | 24 | 26.785 | 1.2672 | 0.1525 |
| 9 | $[4^6 5^6 6^6]$ [c] | 29 | 31.167 | 1.4214 | 0.1260 |
| 10 | $[4^6 6^{12}]$ [d] | 32 | 37.025 | 1.5402 | 0.1067 |

[a] Rhombic dodecahedron
[b] Truncated octahedron
[c] $P6_3/mmc$ $YH_9$ [6]
[d] Irregular chamfered cube, assumed edge asymmetry ratio $r = 1.08$ [5, 8]

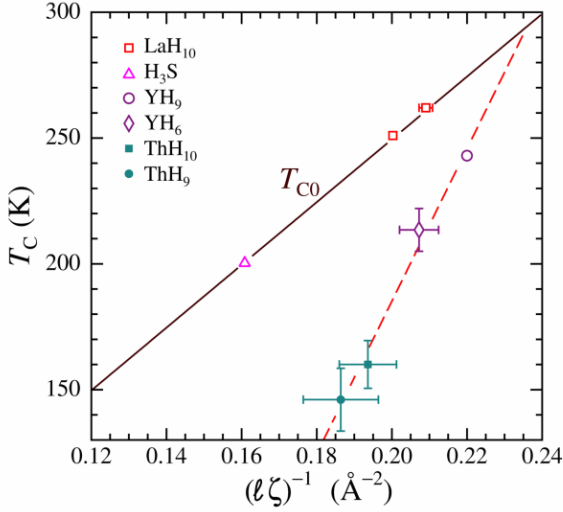

**Fig. 2** Measured transition temperatures $T_C$ for $Fm\bar{3}m$ LaH$_{10}$ [12, 13] and $Im\bar{3}m$ H$_3$S [56] are plotted against $(\ell\zeta)^{-1}$, as determined from the data in Table 1. The solid line represents the interlayer Coulomb pairing theory for $T_{C0}$ defined in Eq. (1). Results for the mixed-phase, non-optimal samples of YH$_9$, YH$_6$, ThH$_{10}$, and ThH$_9$ are also presented for comparison; the red dashed trend line is a fit to these data as described in the text. Horizontal error bars correspond to the lower and upper pressure limits

chamfered cube), containing 14, 24, 29 and 32 vertices, respectively [6, 8].

In the $Fm\bar{3}m$ structure for n = 3, the hydrogens form a rhombic dodecahedron with twelve faces of equal edge length $d$, total surface area $A = 8\cdot 2^{1/2}d^{\,2}$ and distance $\zeta = (6^{1/2}/3)d$ equaling the radius of the inscribed sphere. The [$4^6 6^8$] polyhedron in $Im\bar{3}m$ structure for n = 6 is a truncated octahedron with six square faces of area $A_{sq} = d^{\,2}$, eight hexagon faces of area $A_{hex} = (3\cdot 3^{1/2}/2)d^{\,2}$ and total surface area $A = (6 + 12\cdot 3^{1/2})d^{\,2}$; the distances measured from the polyhedron center to square face centers $\zeta_{sq} = 2^{1/2}d$ and $\zeta_{hex} = (6^{1/2}/2)d$ to hexagon face centers give the area-weighted average $\zeta = (6A_{sq}\zeta_{sq} + 8A_{hex}\zeta_{hex})/A$.

The [$4^6 5^6 6^6$] hydrogen polyhedron for n = 9 in the $P6_3/mmc$ structure comprises six each of quadrangles, pentagons, and hegaxons. Combining theoretical H sites calculated in [6] and measured lattice parameters in [14] for $P6_3/mmc$ YH$_9$, numerical calculations determine $A/d^{\,2} = 31.167$, $\zeta/d = 1.4214$, $C_9 = 0.1260$, and $d/a = 0.3478$, where $d$ is the average of four distinct H-H bond lengths and $a$ is basal lattice parameter. To obtain areas and centroids of the irregular polygons constituting the polyhedral faces, each polygon is divided into triangles based on an edge and an average over polygon edges and vertices is taken. Calculated $A$ and $\zeta$ values are uncertain to $\pm 0.03\%$ and $\pm 0.5\%$, respectively, because of nonplanarity in the quadrangle and hexagon polygons. Nearly the same theoretical $C_9$ values of 0.1263 and 0.1261 are obtained from structural parameter data calculated for CeH$_9$ and ScH$_9$, respectively [6]; $C_9$ values of 0.1260, 0.1272 and 0.1262 are obtained from the structural data reported for synthesized $P6_3/mmc$ phases of CeH$_9$ [62], PrH$_9$ [63], and ThH$_9$ [16], respectively.

The calculations for $Fm\bar{3}m$ YH$_{10}$ repeat the irregular chambered cube model following the theoretical structure determined in [5, 6, 8]. Owing to its tetragonal structure, $I4/mmm$ YH$_4$ does not have a hydrogen clathrate structure per se (sites of 18 hydrogens neighboring an Y may be viewed as occupying the vertices of an elongated dodecahedron [7]) and similarly for hydrogens in orthorhombic $Imm2$ YH$_7$ [15]; from phononic theory, $T_C$ of 87 − 131 and 31 − 46 K are calculated for YH$_4$ and YH$_7$, respectively [15].

### 2.3 Superconducting $M$H$_n$ with $M$ = Y or Th

Other successfully-synthesized hydrogen clathrate compounds, exhibiting superconductivity at high temperatures, include $P6_3/mmc$ YH$_9$, experimentally confirmed to have $T_C$ up to 243 K [14] (theoretically up to 276 K [6]) and $Im\bar{3}m$ YH$_6$ with $T_C$ up to 224 K [14, 15], as well as $Fm\bar{3}m$ ThH$_{10}$ and $P6_3/mmc$ ThH$_9$. Discussions of $Fm\bar{3}m$ YH$_{10}$, which is of special interest because electron-phonon theory predicts $T_C$ above room temperature (303 – 326 K [5, 6, 8, 21]), and CaH$_6$ in the seminal prediction [1] are included in Section 3.1 as potentially superconducting $M$H$_n$ materials.

Of the various YH$_n$ clathrates predicted to be superconductors according to Migdal-Éliashberg theory [5-9, 15, 21, 64], the experimentally verified superconducting syntheses (YH$_3$ is non-superconducting at experimental pressures [14, 65-67]) include yttrium polyhydride mixtures comprising phases of YH$_4$, YH$_6$, YH$_7$, and YH$_9$ [14, 15]. For $P6_3/mmc$ YH$_9$, maximum $T_C$ = 243 K at $P$ = 201 GPa is reported in [14] (sample 2 containing remains of YH$_6$ and traces of YH$_4$) with corresponding lattice parameters $a$ = 3.406 Å and $c$ = 5.210 Å, and mean H-H distance $d$ = 1.185 Å in the theoretical H$_{29}$ polyhedron [6]. Phonon theory predicts $T_C$ = 255 – 276 K at $P$ = 150 GPa [6]. Using C$_9$ = 0.1260 from Table 2 and $\sigma$ = 6, Eq. (6) yields $T_{C0}$ = 274.4 K, which is a hypothetical value in view of predictions of cationic Y [6-8]. Owing to inter-reservoir charge imbalance, $P6_3/mmc$ YH$_9$ is expected to have an intrinsically non-optimal and depressed superconducting transition. A close analogy is found in the compressed $A_3$C$_{60}$ superconductors, where $T_C < T_{C0}$ occurs systematically for alkali constituents $A$ other than pure Cs [27, 41]. The 11% reduction of $T_C$ from $T_{C0}$ for $P6_3/mmc$ YH$_9$ may be interpreted as a pair breaking effect (see, e.g. Eq. (4) in [41]) with pair-breaking energy $\alpha_p$ = 0.12 $k_B T_C$ originating from excess charge in the type I H$_9$ reservoir. Expressing $\alpha_p$ as a fraction $f \approx 0.12$ of Planckian dissipation ($\sim k_B T_C$ at $T_C$), which is related in turn to the excess charge fraction as $f \approx \delta/\sigma$, one obtains from $\delta = \sigma f$ an estimated excess charge of $\sim 0.7e$ per H$_9$. Theoretical charge transfer from Y to the H cage calculated for $P6_3/mmc$ YH$_9$ [6], predicting cationic Y, is evidently contrary the idea of a balanced allocation of 12 valence electrons between Y and H$_9$. Band structure results for YH$_6$, YH$_9$, and YH$_{10}$ in [6] (Fig. S14 caption) also show that H-projected orbitals dominate the density of states near the Fermi level.

In the case of $Im\bar{3}m$ YH$_6$, a range $a$ = 3.529 – 3.582 Å in $bcc$ lattice parameters is measured at pressures from 201 down to 165 GPa in samples containing YH$_6$ [14, 15], corresponding to theoretical average nearest-neighbor H-H distances of $d$ = 1.248 – 1.266 Å between the 24 hydrogens at the vertices of a truncated octahedron structure [6]. The optimal superconducting state requires the charge fraction given by Eq. (3) of $\sigma$ = (6 + 3)/2 = 4.5. From Eq. (6) with $C_6$ = 0.1525 (Table 2), the corresponding calculated optimal $T_{C0}$ is 252 – 265 K. Experimental results for samples denoted as YH$_6$ are $T_C$ = 210 – 227 K (statistically $T_C$ = 219 ± 6 K) [14, 15] and for samples comprising YH$_6$ and other phase material, results are $T_C$ = 214 – 243 K ($T_C$ = 226 ± 12 K). Predictions from Migdal-Éliashberg theory vary over a broad range, $T_C$ = 158 – 301 K for $P$ ranging from 350 down to 100 GPa [6-9, 14] (Section A1). The ~13% discrepancy between $T_{C0}$ predicted by Eq. (6) and measurements of $T_C$ are attributed to several factors. Most fundamental is the expected deviation from charge equilibrium, owing to ionic charge transfer from Y to H$_6$, as deduced from band structure results. Partially displayed PDOS curves for $Im\bar{3}m$ YH$_6$ in [6,8] (limited to $E \geq -20$ eV) show occupation of Y-$d$ orbitals, but these published data are insufficient for determining relative Y and H$_6$ projected charges and whether charge allocation in YH$_6$ is theoretically balanced. However, the findings reported in [7] are that each H atom accepts 0.2$e$, effectively precluding anionic character of the Y ion. The other factor, extrinsic to experiment, is the admixture of phases other than YH$_6$; the samples reported in [14, 15] contain variously $I4/mmm$ YH$_4$, $Im\bar{3}m$ YH$_6$, $Imm2$ YH$_7$, $P1$ Y$_2$H$_{15}$, and an unidentified crystalline form. Multiple phases render the task of ascertaining the actual superconducting component(s) rather ambiguous.

**Table 3** Normalized normal-state-resistance temperature slope $S$ of samples comprising $M$H$_n$ (one or more phases) from published resistance-vs.-temperature curves. Tabulated are reference number, $M$H$_n$ compound(s), sample identification number, reference figure number and curve color, transition temperature $T_C$, pressure $P$, and $S$

| Expt. Ref. | $M$H$_n$ | Sample number | Figure | $T_C$ (K) | $P$ (GPa) | $S$ |
|---|---|---|---|---|---|---|
| [12] | LaH$_{10}$, LaH$_{\sim 12}$ | 3 | 1 – blue | 249 | 159 | 1.08 |
| [12] | LaH$_{10}$, LaH$_{\sim 12}$ | 3 | 2 (a) | 250 | – | 0.91 |
| [12] | LaH$_{10}$ | 1 | 4 – black | 249 | 151 | 0.62 |
| [13] | LaH$_{10}$ | A | 1 – blue | 262 | 188 | 1.06 |
| [13] | LaH$_{10}$ | A | 1 – red | 248 | 196 | 2.8 |
| [15] | YH$_6$, YH$_4$ | K1 | 5 (a) | 224 | 166 | 0.46 |
| [15] | YH$_6$, YH$_7$ | K1 | 5 (b) | 218 | 165 | 0.96 |
| [14] | YH$_6$, YH$_4$ | 2 | 1 (a) – blue | 227 | 201 | 0.37 |
| [14] | YH$_9$, YH$_6$, YH$_4$ | 2 | 1 (a) – red | 243 | 201 | 1.01 |
| [14] | YH$_4$, YH$_6$ | 5 | 1 (d) | 214 | 160 | 0.05 |
| [16] | ThH$_9$ | M2 | 5 (b) | 146 | 170 | 0.39 |
| [16] | ThH$_{10}$ | M3 | 5 (a) | 159 | 174 | 0.18 |
| [16] | ThH$_{10}$, ThH$_9$ |  | S31 (b) | 161 | 170 | 0.34 |
| [56] | H$_3$S | 11/2018 | 1 | 201 | 155 | 0.93 |

For $Fm\bar{3}m$ YH$_3$, which theoretically supports superconductivity [64, 65] but is thus far found to be a normal metal above 5 K [14], integrated electron density analysis indicates acquisition of 1.56$e$ charge on H$_3$ at $P = 120$ GPa (and similarly for $I4/mmm$ YH$_4$) [7]. Adherence to 3$e$ on both Y and H$_3$ would entail hybridization of H 1s$^1$ and Y 4$d$-like orbitals such that the charge on Y is effectively 3$e$.

Of the synthesized ThH$_n$ compounds for n = 4, 6, 9 and 10 reported in [16], the two found to be superconducting are $P6_3/mmc$ ThH$_9$ with $T_C$ of 146 K (170 GPa) and $Fm\bar{3}m$ ThH$_{10}$ with $T_C$ of 159 − 161 K (170 − 174 GPa). In the case of $Fm\bar{3}m$ ThH$_{10}$, the Éliashberg equations are cited [10], predicting $T_C$ of 206 − 227 K at 174 GPa,[4] significantly higher than the measured value. Descriptions of the samples presented in [16] indicate mechanical distortion and inclusion of other Th-hydride phases, including ThH$_4$. Temperature dependences of the samples' resistances shows marked departure from proportionality to $T$ in the normal state, exhibiting instead large extrapolated residual resistance as ratios $R_N(0)/R_N(T_C)$ of 0.61 − 0.82 (Section A2, Table 3), which is a common symptom of non-optimal superconducting material. For example, $H_3S$ prepared without room-temperature annealing exhibits a suppression of $T_C$ (< 200 K) that correlates with increased residual resistance [41]. As superconductors, $ThH_9$ and $ThH_{10}$ may also be intrinsically non-optimal with respect to equilibrium between charges associated with Th and $H_n$. Unfortunately, available information on the electronic structure of $ThH_9$ and $ThH_{10}$ [10, 16] is insufficiently complete for quantifying allocation of the valence charges, although H-projected bands calculated for $Fm\bar{3}m$ $ThH_{10}$ appear to dominate the electronic band structure reported in [10]. Counting the $6d^27s^2$ electron configuration of Th as valence $v = 4$, giving σ = (4 + n)/2, values of hypothetically optimal $T_{C0}$ (> $T_C$) are determined from Eq. (6) and Table 2 using experimentally measured lattice parameters. For $P6_3/mmc$ $ThH_9$, $C_9 = 0.1260$, σ = 6.5, $d = 1.269 – 1.338$ Å, and $T_{C0} = 224 – 249$ K ($P = 86 – 172$ GPa, Table S4 in [16]) and for $Fm\bar{3}m$ $ThH_{10}$, $C_{10} = 0.1067$, σ = 7, $d = 1.185 – 1.231$ Å, and $T_{C0} = 232 – 251$ K ($P = 85 – 183$ GPa, Table S5 in [16]), noting that $d$ relates inversely to $P$.

---

[4] Also presented in [16] are results of calculations for $P6_3/mmc$ ThH$_9$ from the Allen-Dynes formula extrapolated to 170 GPa, predicting $T_C$ of 125 − 147 K. For $Fm\bar{3}m$ ThH$_{10}$, the Allen-Dynes formula yields 160 − 193 K [10, 16].

The above experimental values of $T_C$ for $YH_6$, $YH_9$, $ThH_9$, and $ThH_{10}$ are shown in Fig. 2 as functions of $(\ell\zeta)^{-1} \equiv C_n \sigma^{1/2}/d^2$, where the symbols denote averages and error bars correspond to ranges in $T_C$ and $d$. Results for these non-optimal samples suggest the trend indicated by the dashed line, a linear fit intercepting the solid line for $T_{C0}$ at 293 ± 6 K.

## 3 Discussion

The curious absence of superconductivity in seemingly promising $MH_n$ materials is considered in Section 3.1 from the perspective of charge balance, disorder, and phase purity. The suppressed $T_C$ reported for fcc $LaD_{10}$ in determining the H-D isotope effect [12] is considered in Section 3.2 in terms of materials issues [36], which are shown to be relevant for interpreting the $T_C$ of $D_3S$ [42] and the oxygen isotope effect in high-$T_C$ cuprates [68-70]. A study to detect the signature of the charge transfer potential $e^2/\zeta$ embedded in Eq. (1) by optical reflectivity, similar to its successful use in several high-$T_C$ materials, is proposed for $LaH_{10}$ in Section 3.3, and coexisting electron-phonon and $M$-H electronic Coulomb interactions are contemplated for the case of $LaH_{10}$ in Section 3.4.

### 3.1 Potentially Superconducting Superhydride Clathrates

Virtually all binary superhydrides, except those showing weak coupling [6, 62, 63], are predicted from electron-phonon theory to be superconducting with high transition temperature under high pressure (e.g., [1-10, 15, 20-26]). While a few have been successfully synthesized and shown to be superconducting, an unknown number may have been synthesized but found not to have the predicted high $T_C$ or are non-superconducting. Still others, like $Fm\bar{3}m$ $YH_{10}$, are awaiting to be synthesized [14], or have failed to be synthesized. Accepting that synthesis issues persist in some cases, it is proposed that the superconducting condensate quality in binary clathrates is determined by the chemical details of the enclosed atom and inter-reservoir charge equilibrium. The $Fm\bar{3}m$ binary structure $YH_{10}$ is particularly noteworthy, owing to predictions of $T_C$ exceeding 300 K at high pressures [5, 6, 8, 21], although it has yet to be synthesized [14].

Like its $LaH_{10}$ counterpart, $Fm\bar{3}m$-phase $YH_{10}$ is expected to contain an $H_{32}$ cage in the form of an irregular chamfered cube. Viewing the electronic structure of $LaH_{10}$ as the benchmark model, one may consider that the 3 valence electrons of La, forming La-H hybridized states with the native 10 electrons from H, favorably produce $(13/2)e$ charge in La-projected orbitals and equal charge on $H_{10}$. Equal allocation is enabled by the $4f$ atomic orbitals of La that are otherwise unoccupied. For $YH_{10}$, theoretical PDOS at 250 GPa plotted in Fig. S3(c) of [5] shows occupied $s$, $p$, $d$, and $f$ orbitals projected onto Y, as would be necessary in forming anionic Y. Estimated integrations of PDOS curves over the energy range $-1.5 \text{ Ry} < E < E_F$ indicate electron charges of $5.4e$ in Y orbitals and $6.7e$ in H $s$-$p$ orbitals, which are in the ratio 1.24. The total of $12.1e$, falling short of the expected 13, suggests that $0.9e$ is uncounted, rendering the ratio uncertain. The calculated band structure diagram plotted in Fig. 2 of [8] employs colored tones to show projections onto Y or H states, qualitatively indicating that the charges on $H_{10}$ relative to Y are in ratio of about 1.3. Although charge allocation is not well elucidated by these methods, the band structure results in [5] do find anionic charge on Y and, moreover, that the total charge on $H_{10}$ is just 3% above the optimal value of $6.5e$ corresponding to $\sigma = 6.5$. Calculations at a different pressure could change indicated charge imbalance, although convex hull diagrams of formation enthalpy predict that $YH_{10}$ is stable for $P > 250$ GPa [5] or $P > 225$ GPa [8]. The theoretical uncertainty therefore doesn't rule out synthesizing $YH_{10}$ with $H_{10}$ charge close to the optimal $6.5e$ near the theoretical pressure range of $225 - 300$ GPa for stability [5, 8]. Assuming the theoretical values for lattice parameter $a = 4.600 - 4.747$ Å, average H-H spacing $d = 1.070 - 1.105$ Å, and H-H bond-length ratio $r = 1.09$ [8], application of Eq. (6) and Table 2 evaluates to $T_{C0} = 278 - 296$ K for $Fm\bar{3}m$ $YH_{10}$. Thus, inter-reservoir Coulomb interactions may be considered as a viable mechanism for superconductivity near room temperature in a compressed $YH_{10}$ clathrate.

In the search for high-temperature superconductivity, some researchers have considered other metal-superhydride clathrates as promising candidates, calculating $T_C$ at various high pressures based on strong electron-phonon Éliashberg-based theories [1, 2, 5, 6, 8, 10, 14, 16, 21, 62, 63, 71-73]. Of the superhydrides synthesized with rare earth $M$ [62, 63, 74, 75], no superconductivity for $M$ = Ce has been reported [62, 74], and for $M$ = Pr, emergence of $T_C$ below 9 K is reported [63]. Achieving optimal high-$T_C$ superconductivity, mediated by $M$-H Coulomb interactions in these and other $M\text{H}_n$ clathrates, is contingent on there being an optimal charge allocation between the two charge reservoirs. Verification of Eq. (6) for these other $M\text{H}_n$ clathrates awaits the successful stabilization of optimal superconducting syntheses.

Assuming strong electron-phonon interactions, a sodalite-like $CaH_6$ phase with bcc Ca and truncated octahedron $H_{24}$ cage is predicted from strong-coupled phonon theory to be superconducting in the range $T_C = 220 - 235$ K at a pressure of 150 GPa [1]. More recent studies predict superconductivity in $CaH_n$ with higher H content. Two examples are the $C2/m$ phase $CaH_9$ ($T_C = 240 - 266$ K, $P = 300$ GPa) and $R\bar{3}m$ $CaH_{10}$ ($T_C = 157 - 175$ K, $P = 400$ GPa) [71]. To date, however, no observation of superconductivity in Ca hydrides has been reported and it is uncertain as to whether these materials have even been successfully synthesized. The Ca is evidently a cation in $CaH_6$, owing to a charge transfer to $H_6$ calculated to be $1.02e$ [1]. Integration of the PDOS shown in Fig. S16 in [1] over the range $-20 \text{ eV} \leq E \leq E_F$ (excludes core states for $E < 20$ eV) yields a charge of $2.45e$ on Ca and $5.54e$ on $H_6$. Compared to the optimal $\sigma = 4$ from Eq. (3) with n = 6 and $v = 2$, the $H_6$ reservoir is predicted to be overdoped by $1.54e$. Cation Ca is also predicted for $CaH_9$ and $CaH_{10}$, owing to Ca-to-H charge transfer [71]. Thus, these $CaH_n$ clathrate systems are intrinsically non-optimized, owing to an imbalance of charge between the type II Ca and type I $H_n$ reservoirs, which, with synthesis kinetics and other clathrate system complexities, could explain the absence of a successfully synthesized $CaH_n$ superconductor.

## 3.2 Deuteride Phases: $LaD_{10}$, $D_3S$

Measurement of the H-D isotope effect in $Fm\bar{3}m$ $LaH_{10}$ could provide important information regarding the role played by phonons in the formation of the superconducting state. An H-D mass isotope effect exponent $\alpha = 0.48$ is calculated for $P = 200$ GPa in [23] from the formula for $T_C$ in Allen and Dynes [19] and $\alpha = 0.43$ for $P$ around 160 GPa in [15]. The available data are the resistance transitions shown in Fig. 4 of [12] for $LaH_{10}$ and $LaD_{10}$ samples, showing $T_C = 249$ K at 151 GPa and 180 K at 152 GPa, respectively, seemingly implying $\alpha = 0.46$. Transition midpoint temperatures are lower than $T_C$ by $\Delta T_C \approx 7$ and $\approx 14$ K, respectively. However, the value of $\alpha$ obtained in this manner may not be intrinsic, but rather related to the width of the resistance transition for the $LaD_{10}$ sample since $\alpha \approx 6\ \Delta T_C/T_C$, which curiously has the same factor found for several samples of $D_3S$, wherein $\alpha = (5–7)\ \Delta T_C/T_C$ [42]. Such correlations between mass isotope exponents and widths of superconducting transitions have been observed previously in cuprates [68], where the oxygen mass isotope effect exponent bears the relation $\alpha_O \approx 3\ \Delta T_C/T_C$ for samples of $Y_{1-x}Pr_xBa_2Cu_3O_{7-\delta}$ (here, $\Delta T_C$ is the temperature interval corresponding to 10% and 90% of full transition [68]). Anomalous behavior apparent in $\alpha_O$ is attributed to disorder in non-optimal samples [69, 70]. Taking the various levels of disorder into account, the $T_C$ data for the $D_3S$ and $H_3S$ samples studied in [57] has the corrected exponent value of $\alpha = 0.043 \pm 0.140$ [42], indicating an inconsistency with phonon mediation and necessitating a more unconventional, Coulomb-based approach to the superconductivity in $H_3S$ [41]. The materials issues clearly evident with synthesizing $D_3S$ are not considered in the theoretical treatments regarding the originating superconducting mechanism reviewed in [76]. The observed resistance drop temperatures are deemed inconclusive for an isotope effect that may or may not be present [4].

## 3.3 Interlayer Coulomb Interaction Potential

Optical reflectivity data can provide a direct experimental test for the interlayer Coulombic pairing interaction potential $e^2/\zeta$ imbedded in the expression for $T_{C0}$ in Eq. (1), as discussed in [41] and [49]. In compressed A15-phase $Cs_3C_{60}$ (1.8 GPa pressure), a maximum in the vicinity of 2500 cm$^{-1}$ (0.31 eV) is found in the room-temperature mid-infrared component of the optical conductivity obtained from Kramers-Kronig analysis of ambient-temperature reflectivity (Fig. 1I in [77]), and in compressed $H_3S$ (150 GPa pressure), a dip near 0.46 eV is observed in the normal-state reflectivity (Fig. 4a in [78]). These energies are shown to be proportional to $\zeta^{-1}$ in [41] and can be calculated from the expression [1.0(1) eV Å]$\zeta^{-1}$. A mid-infrared signature near 0.6 eV can, therefore, be projected for $Fm\bar{3}m$ $LaH_{10}$, e.g. from normal-state reflectance measurements, by conjecturing the same $\zeta^{-1}$ scaling with $\zeta$ in the range 1.728 – 1.828 Å determined from cell volume data in [12, 13]. A different optical signature, that of phonons, is proposed in [79] from theoretical temperature dependence of the reflectivity in the superconducting state; it is notably absent above $T_C$.

Given the compelling arguments put forth in [41] comparing A15 $Cs_3C_{60}$ and $Im\bar{3}m$ $H_3S$, measurement and analysis of mid-infrared reflectivity in $LaH_{10}$ (or other hydrogen clathrates) seem an imperative in correctly determining the origin and nature of the pairing mechanism.

### 3.4 Coexisting Phonon/Coulomb Interactions – $LaH_{10}$

Of the recent advances in computational theoretical methods, especially noteworthy are results in [5, 6], which have predicted the synthesis of $Fm\bar{3}m$-phase $LaH_{10}$ [11-13] and also anticipated the superconductivity above 260 K reported in [13]. Calculations of $T_C$ in [5, 6], reaching as high as 286 – 288 K, bear close correspondence to experimental evidence of onset superconductivity at high temperatures in [13, 29]. Combined with subsequent theoretical results in [22-26], phonon mechanisms seemingly appear capable of addressing the full experimental ranges in $T_C$ and $P$ reported in [12, 13], given that the body of wide-ranging theoretical results overlap available data, as illustrated in Section A1 (Figs. 3–4). Data in [13] indicate $T_C \in$ (240, 282) K and occurring within the range $P \in$ (179, 202) GPa, suggesting a strong pressure dependence with mean slope +1.9(6) K/GPa, similar to +2.4(4) K/GPa by fitting data reported in [29] (see Fig. 10 therein). On the other hand, data in [12] comprise $T_C \in$ (243, 251.5) K and $P \in$ (140, 218) GPa forming the "dome" with essentially zero (–0.02 ± 0.02 K/GPa) average slope. The clearly conflicting pressure dependences from these two studies are pointed out in [4] as suggestive of different phases (discrepancies in phase and stoichiometry of the samples studied in [12, 13] do indicate the presence of non-optimal material). However, there is no conflict related to the optimal phase found from each study, where the results for optimal $T_C$ and $P$ (Table 1) are connected by the same

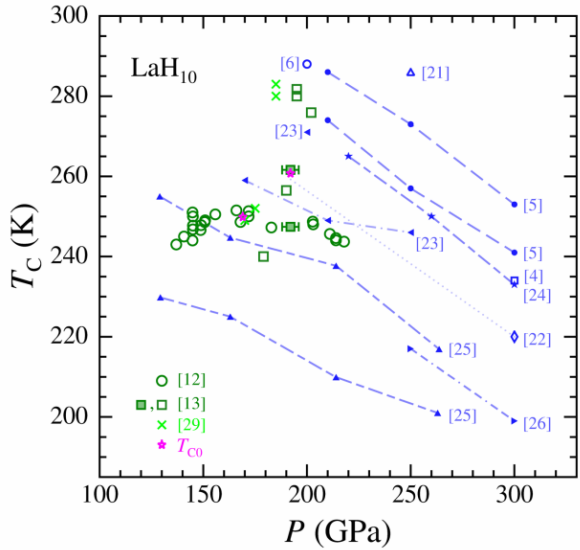

**Fig. 3** Plot of transition temperature $T_C$ vs. pressure $P$ for $Fm\bar{3}m$ $LaH_{10}$ (marked with source references). Larger green symbols denote experimental data from [12, 13, 29]; smaller blue symbols and connecting lines represent theoretical results from [4-6, 21-26]. Magenta star symbols are at $T_{C0}$ and $P$ as given in Table 1

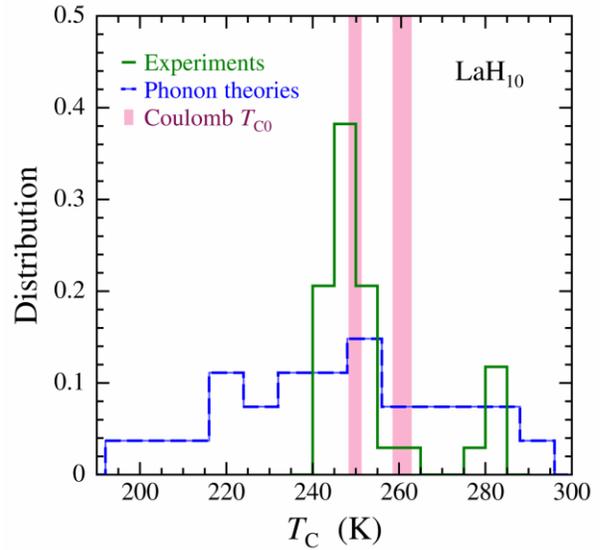

**Fig. 4** Distributions of experimental and theoretical results for $T_C$ of $Fm\bar{3}m$ $LaH_{10}$, as presented in Fig. 3 and Table 1. Normalized histograms show distributions in $T_C$ from experiments (solid green) and phonon theories (dashed blue). Vertical pink bars denote $T_{C0}$ presented in Table 1 (bar width denotes uncertainty)

La-H electronic Coulomb interaction (Eq. (1) and Fig. 2).

The combination of the dissonant phonon-based theoretical $T_C(P)$ results with the accordant Coulombic-based findings of Eq. (1) for optimal $T_C = T_{C0}$ augurs well for a comprehensively applicable treatment, even though various mathematical approximations, assumptions, and rigor have thus far been limited to essentially conventional electron-phonon calculations of $T_C$ [4-6, 22-26]. To this end, the interplay between the inter-reservoir Coulomb interactions underlying Eq. (1) with phononic mechanisms may be considered. Non-optimal superconducting phases are manifested mainly as different temperatures relative to the optimal $T_C$ in [13] and by different pressures relative the optimal $P$ in [12]. Optimal $T_C$, therefore, denotes a superconducting-phase fixed point that is determined by the electronic La-H interactions. Shallowness of the $T_C$-vs.-$P$ dome in LaH$_{10}$ (< 4% variation in $T_C$) suggests that superconductivity in the non-optimal material may be complimented by sufficiently strong electron-phonon coupling [25], given that variations of $T_C$ with pressure in A15-phase Cs$_3$C$_{60}$ [27] or doping in cuprate superconductors [31] are comparatively larger by an order of magnitude. Coexistence of Coulombic and phononic mechanisms in non-optimal material might be observable. Possible evidence to consider is the broad resistance drop near 225 K for sample 3 of LaH$_{10}$ in [12] (Fig. 1, blue curve), where $T_C$ = 249 K and $P$ = 150 GPa are both less than the optimal values in Table 1. It is also worth noting that a possible source of at least some of the phonon-related phenomena in LaH$_{10}$ may be a sympathetic response of the lattice to the formation of the Coulomb-mediated superconducting state, as suggested for Cs$_3$C$_{60}$ [49]. At present, more definitive evaluations of electronic Coulomb interactions, electron-phonon mechanisms, and their interplay await more extensive experimental study. Contributions of interlayer Coulomb interactions to conventional electron-phonon superconductivity are treated for electronically layered structures in [80].

## 4 Conclusion

Available experimental data on LaH$_{10}$ and YH$_9$ samples contain persuasive evidence of unconventional high-$T_C$ superconductivity in these compounds, including a maximum in $T_C(P)$ [12, 14], suggesting optimization, and a linear temperature dependence in the normal-state resistance [30], which places upper bounds on the electron-phonon coupling strength, owing to the absence of resistance saturation [43]. Although theory based on phonon mediation notably predicted superconductivity at high temperatures [5, 6], convergence of broad ranges in calculated $T_C$ to experimental values is hampered by theoretical pressure dependences [5, 24-26] that are contrary to experiment [29], thus necessitating an alternate approach.

Examination of theoretical band structure shows that $\sigma = (v + n)/2$ holds for $Fm\bar{3}m$ LaH$_{10}$, owing to the anionic nature of the La ion [22]. However, theory for $P6_3/mmc$ YH$_9$ [6] indicates that cationic Y forms H$_9$-related charges exceeding the optimal $\sigma = 6$, and similarly for $Im\bar{3}m$ YH$_6$ and $Fm\bar{3}m$ YH$_{10}$ [6-8]. When optimal charge equilibrium is intrinsically unrealizable, the experimental $T_C$ is expected to be suppressed relative to $T_{C0}$. For $Fm\bar{3}m$ LaH$_{10}$, optimal transition points $T_C$ of 251(1) K and 262(1) K at optimal pressures of 169(4) and 192(4) GPa are identified from results in [12] and [13], respectively. The corresponding calculated $T_{C0}$ values are 249.8(1.3) and 260.7(2.0) K, respectively, which compare extremely well (with a < 2 K uncertainty) with experimental values. For $P6_3/mmc$ YH$_9$, the highest experimental $T_C$ of 243 K in a mixed-phase sample is depressed by 11%, attributable to an overdoped H-cage; the phononic theory prediction is 255 − 276 K [6]. Transition temperatures reported for nominally YH$_6$ mixed-phase samples [14, 15], and for ThH$_9$ and ThH$_{10}$ [16] samples with large residual resistances, fall short for either optimal Coulomb interactions or phonon interactions. Comparison of theory to experiment is, unfortunately, clouded by the ambiguities of interpreting mixed phase samples and suppression of $T_C$ by intrinsic deviation from $M$-H$_n$ charge equilibrium, found theoretically for $M$ = Y, and extrinsic effects of sample disorder, notably for $M$ = Th.

Mass isotope data for DH$_{10}$ reported in [12], deemed inconclusive in [4], are shown to more likely reflect the inverse correlation between $T_C$ and the fractional transition width $\Delta T_C/T_C$, as is found for D$_3$S. Consequently, and notwithstanding the predictions of Éliashberg theory, the precise role played by phonons in the superconductivity of $M$H$_n$ awaits further investigation. Since quantitative scaling with the theoretical $e^2/\zeta$ is noted from optical evidence for compressed A15 Cs$_3$C$_{60}$ [49] and H$_3$S [41], and plausibly extending to LaH$_{10}$, reflectivity measurements of $M$H$_n$ may prove invaluable in determining the nature of the mediating boson.

**Acknowledgments** The authors are grateful for support from the College of William and Mary, New Jersey Institute of Technology, and the University of Notre Dame. We also acknowledge helpful information provided by Dr. F. Peng.

**Funding Information** This study was supported by Physikon Research Corporation (Project No. PL-206) and the New Jersey Institute of Technology.

**Compliance with Ethical Standards**

**Appendix**

The following appendix sections present and analyze results for the pressure variation in $T_C$, normal-state resistance, pressure dependence of the lattice parameter, and accuracy and systematic error of the model, based on this work and published data.

**A1 Pressure Variation in $T_C$**

A comprehensive plot of available experimental data [12, 13, 29] and theoretical results [4-6, 21-26] pertaining to the variation of $T_C$ with pressure $P$ in LaH$_{10}$ is presented in Fig. 3. Open circle symbols shown for $P \in$ (140, 218) GPa and $T_C \in$ (243, 251) K are transcriptions of experimental data from Fig. 1 inset of [12] (error bars omitted for clarity). The arched trend in $T_C$ vs. $P$, noted in [12], comprises points with highest $T_C$ at 251.5 and 251.3 K (error bars ± 1 K) corresponding to $P$ at 166 and 172 GPa (± 2 GPa), respectively (data for sample 5 in [12]). Taking rounded averages and uncertainties, the values 251(1) K and 169(4) GPa thus determine the experimentally measured optimal $T_C$ and $P$, respectively, shown in the first row of Table 1.

Although experimental measurements of $T_C$ are not specified in [13], full superconducting resistance transition curves for cooling and warming of sample A are presented in Fig. 3 therein. Onset values of $T_C$ are determined herein to about ± 1 K uncertainty from the intersection of tangents drawn through the linear-in-T normal-state region (see Section A2) and the more steeply sloped region just below the transition onset. This method yields $T_C$ of 262(1) and 247(1) K for cooling and warming, respectively, and is shown as the filled square symbols in Fig. 3, where horizontal error bars span the pressures of 188 and 196 GPa associated with the data (i.e. $P = 192 \pm 4$ GPa). The higher $T_C$ for cooling is assumed to be closer to optimal and therefore entered in the second row of Table 1. Coincidentally, Fig. 3 shows by way of comparison that the onset $T_C$ for warming lies in the region of the lower $T_C$ observed in the data from [12] (sample 6 therein). Other resistance-*vs*.-temperature data in [13] show resistance drops suggesting onsets of superconducting behavior (full transitions not displayed). Figure 4 in [13] shows a resistance drop at 280 K for 0.1 mA applied current and 195 GPa pressure for sample F. Figure S4 in [13] shows drops at 257, 282, and 276 K for pressures of 190, 195, and 202 GPa, respectively, for sample B. Additionally, a small jump at 240 K (at assumed 179 GPa) is described for sample G [13]. The five open square symbols in Fig. 3 represent these other temperatures and pressures drawn from [13]. Cross symbols represent data from Fig. 10 of [29].

Small symbols in Fig. 3 display results of theoretical calculations of $T_C$ for LaH$_{10}$ from strong electron-phonon coupling based on Migdal-Éliashberg theory, density functional theory or *ab initio* methods (labels distinguish the source references [4-6, 21-26], exclusive of ranges in theoretical stability study of [81]). Results calculated at multiple values of $P$, connected by broken lines, show that the electron-phonon mechanism reproducibly predicts that $T_C$ has a monotonically decreasing dependence on $P$ with average slope $\Delta T_C/\Delta P \approx -0.30 \pm 0.08$ K/GPa [5, 23-26], a finding with theoretical basis [26]. The dotted line in Fig. 3 illustrates the linear extrapolation from the calculated point (diamond symbol) to the range of experimental pressure that is used in [22]. Superconducting density functional theory (SCDFT), as applied in [23], predicts $T_C = 271$ K at $P = 200$ GPa, whereas SCDFT applied in [25] predicts a much lower $T_C = 214$ K (interpolated to $P = 200$ GPa).

The magenta star symbols in Fig. 3 are the results for $T_{C0}$ given in Table 1 at the values of $P$ indicated. These two points define a positive change $\Delta T_C/\Delta P = +0.47$ K/GPa, in contrast to negative variations predicted in electron-phonon theory [5, 22, 24-26].

The distributions of the $T_C$ values from Fig. 3 are cast in Fig. 4 in the form of histograms corresponding to the experiments (solid green) and phonon theories (dashed blue) normalized to the number of points (30 and 27, respectively), and with bin sizes of 5 K and 10 K, respectively. Statistical averages and $\pm$ standard deviations of the distributions are as follows: $T_C = 251 \pm 10$ K (experiments) and $245 \pm 26$ K (phonon theories); $P = 174 \pm 27$ GPa (experiments) and $232 \pm 52$ GPa (phonon theories). Theoretical pressure dependence suggests shifting the statistical

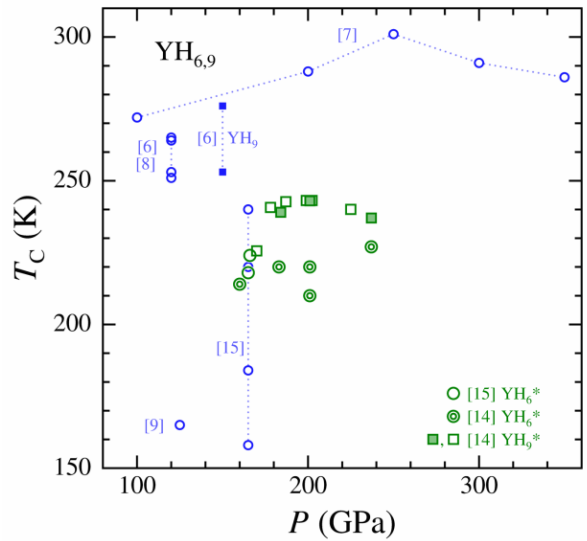

**Fig. 5** Plot of transition temperature $T_C$ *vs*. pressure $P$ for $Im\bar{3}m$ YH$_6$ and $P6_3/mmc$ YH$_9$ (marked with source references), denoted by circular and square symbols, respectively. Larger green symbols denote experimental data [14, 15]; smaller blue symbols and connecting lines represent phonon theories [6-9, 15]

average for the phonon theories as $T_C = 263 \pm 29$ K at $P \to 174$ GPa. Pre-experimental theories include predictions of 274 and 286 K at 210 GPa in [5] and 288 K at 200 GPa in [6], which may be compared to resistance drops [13] or magnetic onsets [29] at temperatures of 276 – 283 K and pressures of 185 – 202 GPa, and represented by the smaller histogram fragment in Fig. 4. The two pink vertical bars (and widths) in Fig. 4 represent the values of $T_{C0}$ (and uncertainties) given in Table 1.

Figure 5 is a comprehensive plot of $T_C$ vs. $P$ for $YH_6$ (circle symbols) and $YH_9$ (square symbols) where the larger green circles and squares denote experimental data in [14, 15]. The smaller blue circles and filled squares denote theoretical results for $YH_6$ [6-9, 14] and $YH_9$ [6], respectively. Asterisk symbols in the legend express the caveat that phases other than those indicated are determined from X-ray data analysis (e.g. $YH_4$) [14]. Statistical averages and ± standard deviations of samples identified as $YH_6$ in [14, 15] are $T_C = 219 \pm 5$ K and $P = 188 \pm 28$ GPa; statistics for the $YH_6$ phonon theories are $T_C = 245 \pm 47$ K and $P = 176 \pm 75$ GPa. The "dome" trend in $T_C$ vs. $P$ for $YH_9$ with maximum $T_C = 243$ K at $P = 201$ GPa is noted in [14] (filled symbols, before change in $P$; open symbols, after change in $P$); phonon theory predicts $T_C = 253 - 287$ at $P = 150$ GPa [6].

## A2 Normal State Resistance

Unique features signifying unconventional high-$T_C$ superconductivity are a linear temperature dependence in the normal state resistivity, $\rho_N(T) \propto T$, and the absence of saturation at high temperature, such that the negative curvature of $\rho_N(T)$ for conventional strong electron-phonon coupled metals is absent [43]. Such unconventional behavior in $\rho_N(T)$, observed in optimally- and over-doped high-$T_C$ cuprates, is ascribed generally to "Planckian" dissipation phenomena [44, 45]. Superconducting $LaH_{10}$ and $YH_9$ are exemplary of this behavior for $T > T_C$, as can be seen in Fig. 6 showing resistance vs. temperature for (a) $Fm\bar{3}m$ $LaH_{10}$ sample 3 from Fig. 1 of [12] (blue curve, $P = 150$ GPa, $T_C \sim 249$ K) and (b) $P6_3/mmc$ $YH_9$ sample 2 from Fig. 1a of [14] (red curve, $P = 201$ GPa, $T_C = 243$ K). Dotted lines illustrate linear dependence for $T > T_C$ extrapolating to the origin. Resistance of $LaH_{10}$ sample A in [13] (Fig. 3 therein, cooling curve) and $Im\bar{3}m$ $YH_6$ sample M1 in [15] are also consistent with normal-state linearity in $T$. Table 3 summarizes normalized normal-state slopes obtained from the temperature dependence of the resistance $R_N(T)$ reported for various samples of La, Y, and Th superhydrides and $H_3S$, defined as $S = [T_C/R_N(T_C)] (\Delta R_N/\Delta T)$ in terms of slope $\Delta R_N/\Delta T$ taken asymptotically

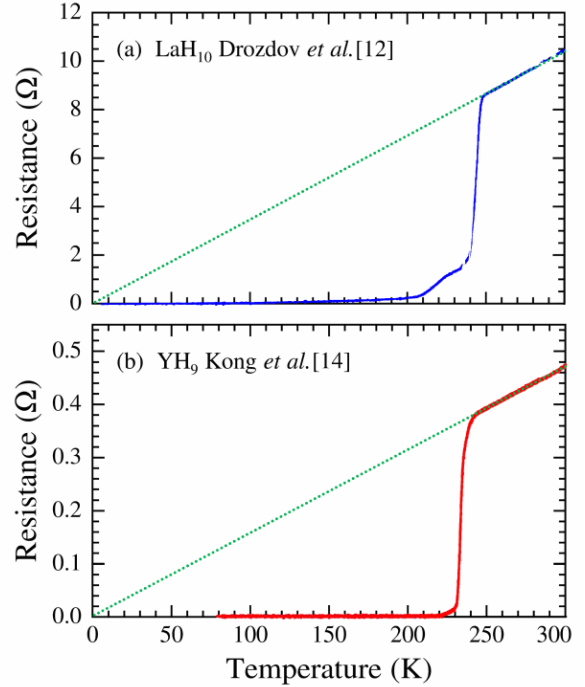

**Fig. 6** Resistance vs. temperature for (a) $Fm\bar{3}m$ $LaH_{10}$ sample 3 from Fig. 1 of [12] (blue curve) and (b) $P6_3/mmc$ $YH_9$ sample 2 from Fig. 1(a) of [14] (red curve). Dotted lines show extrapolations of normal-state resistance to the origin

above $T_C$ and $R_N(T_C)$ as extrapolated to $T_C$. Resistance in applied magnetic fields, which depress the superconducting transition, shows linear-in-$T$ extending to $T < T_C$; $S = 0.91$ at 9 Tesla is obtained for sample 3 from Fig. 2 (a) in [12] and extension to above 60 Tesla is reported in [30]. Extrapolated residual resistance ratio is given by $R_N(0)/R_N(T_C) = 1 - S$. The statistical distribution in $S$ with bin size 0.25 is shown in Fig. 7 (from Table 3 and near unity $S$ assumed from [30]), indicating that highest probability occurs at $S \approx 1$. Ideally, $S$ approaches unity for (unconventional) optimal high-$T_C$ superconductors.

Dissipation from inelastic electron scattering in the normal state is determined from the scattering rate formula $\tau_N^{-1} = \rho_N e^2 n/m$, written in terms of density $n$ and mass $m$ of the carriers. An estimated $\rho_N \sim 10^{-5}$ $\Omega$ cm at $T \approx 285$ K is calculated for sample F in [13] from measured resistance $\approx$ 0.11 m$\Omega$ at 0.1 mA indicated in Fig. 4, thickness $\sim$ 2 $\mu$m mentioned in the supplemental material of [13], and an assumed Van der Pauw factor of 4.53. Carrier density $n$ is obtained from the charge fraction $\sigma = 6.5$ associated with $H_{10}$ (6.5$e$ per formula unit volume), $m$ is derived from theoretical superconducting parameters as $m = \hbar k_F/v_F = \hbar^2 k_F/\pi\Delta\xi$ with theoretical gap $\Delta = 2.3\ k_B T_C$ [22] and experimental coherence distance $\xi = 1.7$ nm (average determined in [12]); Fermi wave vector is determined as $k_F = (3\pi^2 n/g)^{1/3}$ with $g = 2$ (approximating the theoretical Fermi surface in [22] as two equal-area spheres). The Planckian dissipation energy evaluates to $\hbar\tau_N^{-1} \sim 4k_B T$ from the above data for LaH$_{10}$. Acknowledging that this analysis provides only a single order of magnitude estimation, the dissipation may be considered as similar to $(1.3 - 2.3)k_B T$ determined for optimal cuprate superconductors [46-48]. While conventional theory of strong electron-phonon scattering might mimic linearity for $T$ between $T_C$ and 300 K, if designed with judicious choices of residual and saturation resistivities, it appears rather implausible in view of the preponderance of $S \approx 1$ listed in Table 3 and displayed in Fig. 7.

### A3 Pressure Dependence of the Lattice Parameter

X-ray diffraction is employed to determine volumes of LaH$_{10}$ referenced to the fcc cubic cell for samples 2 and 3 in [12] and referenced per La for superconducting samples A–C and E–G in [13]. Samples 2 and 3 are reported to contain minority phase material [12]. Samples F and G are described as mixed phase [13]. Pressure dependence in the fcc lattice parameter $a$ are plotted in Fig. 8 as filled circles (data from [12]) and filled squares ([13]). Square symbols inscribed with a white cross correspond to samples C and E for which no information on transition temperature is provided in [13]. The straight line is a linear fit (exclusive of the two white-crossed squares) to the relation $a = a_0 - sP$

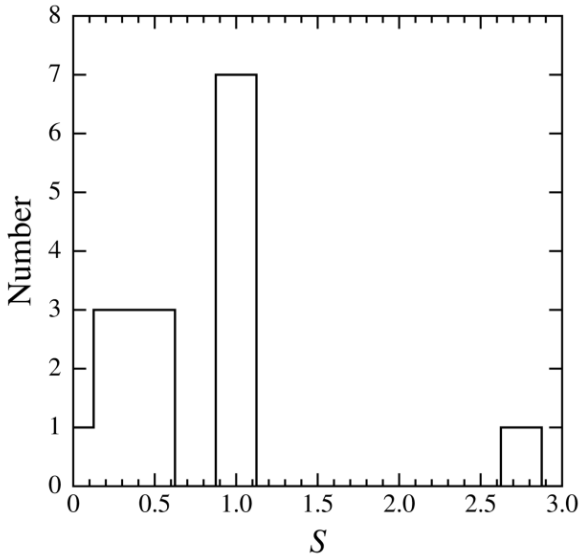

**Fig. 7** Statistical distribution of data for normalized normal-state resistance temperature slope parameter $S$ from Table 3

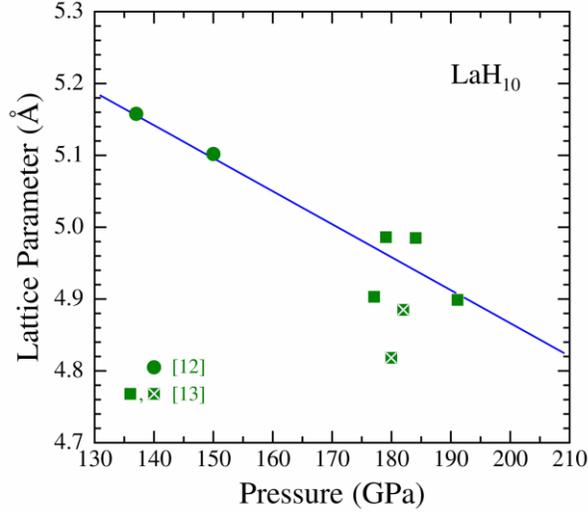

**Fig. 8** Lattice parameter (face centered cubic) *vs.* pressure for $Fm\bar{3}m$ LaH$_{10}$ from X-ray data in [12] (circles) and [13] (squares). The solid line shows the linear dependence obtained by fitting the data exclusive of the two white-crossed squares (see text)

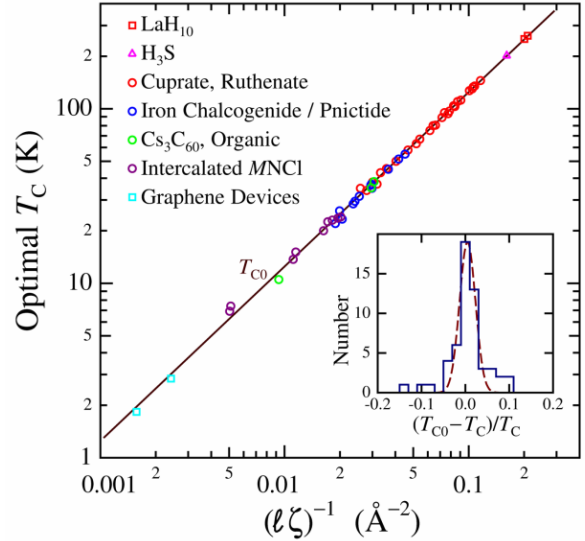

**Fig. 9** Optimal measured transition temperatures $T_C$ *vs.* $(\ell\zeta)^{-1}$ for 55 superconductors (grouped as per legend) are shown in comparison with calculated $T_{C0}$ (solid line). Inset: statistical distribution of fractional deviation between calculated $T_{C0}$ and measured $T_C$; dashed curve shows fitted Gaussian form

with intercept $a_0 = 5.78 \pm 0.15$ Å, slope $s = 0.00459 \pm 0.000091$ Å/GPa, and root-mean-square deviation of 0.043 Å between the data and the line. The fitted function is used to extract values of $a$, and therefrom $d$, $A$, $\zeta$, $\ell$, and $T_{C0}$, at experimental optimal pressures (Table 1).

## A4 Accuracy and Systematic Error of the Model

The accuracy of Eq. (1) rests on that of the constant $\Lambda$, which was originally obtained by fitting $\Lambda$ in the right-hand side expression to experimental data for $T_{C0}$ taken for the left-hand side. The fitted data included 31 phase-pure, optimal high-$T_C$ compounds, with superconducting transitions in the range 10.5 K $\leq T_C \leq$ 145 K [40]. Subsequently, 24 other compounds (see [41, 49-55]), optimized according to the developed criteria and inclusive of the two LaH$_{10}$ samples, have been added to the list of optimal superconductors with $T_C$ values demonstrated to be in congruence with Eq. (1), bringing the total number to 55 from eleven disparate superconducting families (with $T_C$ ranging from ~2 to 260 K). Comparisons of $T_{C0}$ calculated for LaH$_{10}$ at 169(4) GPa [12] and at 192(4) GPa [13] with the measured average $T_C$ values (from Table 1) yield differential accuracies of 0.48% and 0.50%, respectively. The optimal transition temperatures are predicted with an overall statistical accuracy of $\pm$ 4.3% over a range in $T_C$ from ~2 to 260 K, with an average fractional deviation of 0.7%. These results are graphically represented in Fig. 9, which plots the measured optimal transition temperature $T_C$ *vs.* $(\ell\zeta)^{-1}$, and compared to the theoretical expression for $T_{C0}$ given in Eq. (1) represented by the solid line. Deviations of $T_C$ from the theoretical line have a root-mean-square average of 1.3 K. The statistical distribution of fractional deviation between calculated $T_{C0}$ and measured $T_C$ is given in the inset of Fig. 9, where the dashed curve shows the fitted Gaussian form.

Nearly abrupt superconducting transitions are also predicted theoretically for the hydrides LaH$_{10}$ and H$_3$S, by virtue of their being three-dimensional bulk crystals. One might assume that improved synthesis of the hydride samples would lead to sharper transitions, reduced uncertainty in determining $T_C$ and, in turn, the value of $\Lambda$.